\documentclass[aps,preprint,preprintnumbers,amsmath,epsf,showpacs,amssymb,nofootinbib]{revtex4}
\usepackage{amssymb}
\usepackage{amsmath}
\usepackage{bm}
\usepackage{graphicx}
\draft
\begin{document}
\title{Primordial dark energy from a condensate of spinors in a 5D vacuum}
\author{$^{1, 2}$ Pablo Alejandro S\'anchez\footnote{pabsan@mdp.edu.ar}, and $^{1, 2}$Mauricio
Bellini\footnote{mbellini@mdp.edu.ar}}
\address{$^{1}$ Departamento de F\'{\i}sica, Facultad de Ciencias Exactas y
Naturales, \\
Universidad Nacional de Mar del Plata, Funes 3350, (7600) Mar del
Plata, Argentina. \\ \\
$^{2}$ Instituto de Investigaciones F\'{\i}sicas de Mar del Plata
(IFIMAR), Consejo Nacional de Investigaciones Cient\'{\i}ficas y
T\'ecnicas (CONICET), Argentina.}
%%%%%%%%%%%%%%%%%%%%%%%%%%%%%%%%%%%%%%%%%%%%%%%%%%%%%%%%%%%%%%%%%%%%%%%%%%%%%%%%%%%%%%%%%%%%%%%%%%%%%%%%%%%%%%%%%%%%%%%%%%%%
\begin{abstract}
We explore the possibility that the expansion of the universe can
be driven by a condensate of spinors which are free of
interactions on a 5D relativistic vacuum defined on an extended de
Sitter spacetime which is Riemann-flat. The extra coordinate is
considered as noncompact. After making a static foliation on the
extra coordinate, we obtain an effective 4D (inflationary) de
Sitter expansion which describes an inflationary universe. We
found that the condensate of spinors here studied could be an
interesting candidate to explain the presence of dark energy in
the early universe. The dark energy density which we are talking about is
poured into smaller sub-horizon scales with the evolution of the inflationary expansion.
\end{abstract}
%%%%%%%%%%%%%%%%%%%%%%%%%%%%%%%%%%%%%%%%%%%%%%%%%%%%%%%%%%%%%%%%%%%%%%%%%%%%%%%%%%%%%%%%%%%%%%%%%%%%%%%%%%%%%%%%%%%%%%%%%%%%
\maketitle

\section{Introduction}

Modern versions of 5D General Relativity abandon the cylinder and
compactification conditions used in original Kaluza-Klein (KK)
theories, which caused problems with the cosmological constant and
the masses of particles, and consider a large extra dimension. The
main question that these approaches address is whether the
four-dimensional properties of matter can be viewed as being
purely geometrical in origin. In particular, the Induced Matter
Theory (IMT)\cite{IMT} is based on the assumption that ordinary
matter and physical fields that we can observe in our 4D universe
can be geometrically induced from a 5D Ricci-flat metric with a
space-like noncompact extra dimension on which we define a
physical vacuum. The Campbell-Magaard Theorem (CMT)\cite{campbell}
serves as a ladder to go between manifolds whose dimensionality
differs by one. Due to this theorem one can say that every
solution of the 4D Einstein equations with arbitrary energy
momentum tensor can be embedded, at least locally, in a solution
of the 5D Einstein field equations in a relativistic vacuum:
$G_{AB}=0$\footnote{We shall consider that capital letters $A,B$ run from
$0$ to $4$ in an 5D extended de Sitter spacetime (where the 3D Euclidean space is in cartasian coordinates),
small letters $a,b$ run from $0$ to $5$ in
a 5D Minkowsky spacetime (in cartasian coordinates), Greek letters $\alpha,\beta$
run from $0$ to $3$ and latin letters $i, j$ run from $1$ to $3$.}. Due to this
fact the stress-energy may be a 4D manifestation of the embedding
geometry and therefore, by making a static foliation on the
space-like extra coordinate of an extended 5D de Sitter spacetime,
it is possible to obtain an effective 4D universe that suffered an
exponential accelerated expansion driven by an effective scalar
field with an equation of state typically dominated by
vacuum\cite{ua,ua1,LB,B}. An interesting problem in modern
cosmology relies to explain the physical origin of the
cosmological constant, which is responsible for the exponential
expansion of the early inflationary universe. The standard
explanation for the early universe expansion is that it is driven
by the inflaton field\cite{guth}. Many cosmologists mean that such
acceleration (as well as the present day accelerated expansion of
the universe) could be driven by some exotic energy called {\em
dark energy}. Most versions of inflationary cosmology require of
one scalar inflaton field which drives the accelerated expansion
of the early universe with an equation of state governed by the
vacuum\cite{1}. The parameters of this scalar field must be rather
finely tuned in order to allow adequate inflation and an
acceptable magnitude for density perturbations. The need for this
field is one of the less satisfactory features of inflationary
models. Consequently, we believe that it is of interest to explore
variations of inflation in which the role of the scalar field is
played by some other field\cite{2,3}. Recently has been explored
the possibility that such expansion can be explained by a
condensate of dark spinors\cite{Lee}. This interesting idea was
recently revived in the framework of the Induced Matter Theory
(IMT)\cite{premio}. In this work we shall extend this idea.

\section{The Effective Lagrangian in 5D Riemann-flat Spacetime}

We are concerned with a 5D Riemann-flat spacetime with a line
element given by:
    \begin{equation}\label{met1}
        dS^{2}=\left(\frac{\psi}{\psi_{0}}\right)^{2}\left[dt^{2}-e^{\frac{2t}{\psi_{0}}}(dx^{2}+dy^{2}+dz^{2})\right]-d\psi^{2},
    \end{equation}
where ${t, x, y, z}$ are the usual local spacetime coordinate system and $\psi$ is the noncompact space-like extra dimension.
%%%%%%%%%%%%%%%%%%%%%%%%%%%%%%%%%%%%%%%%%%%%%%%%%%%%%%%%%%%%%%%%%%%%%%%%%%%%%%%%%%%%%%%%%%%%%%%%%%%%%%%%%%%%%%%%%%%%%%%%%%%%

We start from an effective Lagrangian density for non massive
fermions in 5D:
    \begin{equation}
        \mathcal{L}_{eff}=-\frac{1}{2}(\nabla_{A}\overline{\Psi})(\nabla^{A}\Psi).
    \end{equation}
At this point it is easy to obtain the equations of motion from a variational principle. The Euler-Lagrange
equations for both $\Psi$ and $\overline{\Psi}$ can be obtained making the functional derivatives:
    \begin{eqnarray}
        \frac{\delta\mathcal{L}_{eff}}{\delta\overline{\Psi}}&=&0,
        \\
        \nabla_{A}\frac{\delta\mathcal{L}_{eff}}{\delta(\nabla_{A}\overline{\Psi})}&=&\frac{1}{2}\nabla_{A}\nabla^{A}\Psi, \\
        \frac{\delta\mathcal{L}_{eff}}{\delta g_{MN}}  & = & \frac{1}{2} \nabla^M \overline{\Psi} \nabla^N \Psi + \nabla_P J^{MNP},
    \end{eqnarray}
where the effective current $J^{MNP}$ is symmetric with respect to permutations of $M$ and $N$
\begin{equation}
J^{MNP} = \frac{1}{8} \left(\nabla^M \overline{\Psi} f^{NP} \Psi + \overline{\Psi} \nabla^M \Psi\right),
\end{equation}
and $f^{NP} = \left[\gamma^N,\gamma^P\right]$ is antisymmetric\cite{Bohmer}.
At this point we are in conditions of introduce the stress tensor $T^{MN} =2\frac{\delta\mathcal{L}_{eff}}{\delta g_{MN}}-
g^{MN} \,{\mathcal L}_{eff}$
\begin{equation}
T^{MN} =  \nabla^M \overline{\Psi} \nabla^N \Psi + 2\nabla_P J^{MNP} +  \frac{1}{2} g^{MN} \left( g_{AB}\,
\nabla^A \overline{\Psi} \nabla^B \Psi \right).
\end{equation}

Applying the compatibility condition on the metric
$\nabla_{C}g^{AB}=0$, we obtain
    \begin{equation}
        \nabla_{A}\nabla^{A}\Psi=\nabla_{A}(g^{AB}\nabla_{B}\Psi)=(\nabla_{A}g^{AB})\nabla_{B}\Psi+g^{AB}\nabla_{A}\nabla_{B}\Psi=0,\nonumber
    \end{equation}
we obtain that the equation for the spinor $\Psi$ takes the form
    \begin{equation}\label{mot}
        g^{AB}\nabla_{A}\nabla_{B}\Psi=0.
    \end{equation}
The same procedure yields an identical equation for the field $\overline{\Psi}$ .
On the other hand, the 5D Einstein equations for the Riemann-flat metric (\ref{met1}), is
\begin{equation}
\left<0\left| T_{AB} \right|0\right> =0,
\end{equation}
where $\left<0\left|T_{AB} \right|0\right>$ denotes the
expectation value of $T_{AB}$ in the vacuum state $|0>$.

\subsection{Tensorial Formulation of the Equation of Motion}

Using the formalism previously introduced, the double-Nabla can be expressed explicitly:
    \begin{eqnarray}
        \nabla_{A}\nabla_{B}\Psi &=& \partial_{A}\nabla_{B}\Psi+\Gamma_{A}\nabla_{B}\Psi-\omega_{AB}^{\enskip\enskip C}\nabla_{C}\Psi\nonumber
        \\
        &=& \partial_{A}\partial_{B}\Psi-\frac{1}{4}\partial_{A}\omega_{B}^{\enskip ab}\gamma_{a}\gamma_{b}\Psi-
        \frac{1}{4}\omega_{B}^{\enskip ab}\gamma_{a}\gamma_{b}\partial_{A}\Psi-\frac{1}{4}\omega_{A}^{\enskip ab}\gamma_{a}\gamma_{b}\partial_{B}\Psi\nonumber
        \\
        & & +\frac{1}{16}\omega_{A}^{\enskip ab}\omega_{B}^{\enskip cd}\gamma_{a}\gamma_{b}\gamma_{c}\gamma_{d}\Psi-\omega_{AB}^{\enskip\enskip
        C}\partial_{C}\Psi+
        \frac{1}{4}\omega_{AB}^{\enskip\enskip C}\omega_{C}^{\enskip ab}\gamma_{a}\gamma_{b}\Psi,\nonumber
    \end{eqnarray}
where the spin connection is $\omega_{M}^{\enskip
ab}=-e_N^{\enskip a} \left[
\partial_M e_A^{\enskip b} g^{AN} + e_B^{\enskip b} g^{AB}
\Gamma^N_{\enskip AM} \right]$ and $\Gamma_M = -{1\over 8} \omega_M^{\,\,ab} \left[\gamma_a, \gamma_b\right]$
\footnote{The tensors can be written using the vielbein $e^A_{\,\,a}$
and its inverse ${e}^a_{\,\,A}$, such that, if $e^A_{\,\,a}
{e}^b_{\,\,A} = \delta^b_a$ and
\begin{equation}
\eta_{ab} = e^A_{\,\,a}e^B_{\,\,b} g_{AB},
\end{equation}
where $\eta_{ab}$ is the 5D Minkowsky tensor metric with signature $(+,-,-,-,-)$.}.

Thus, after replacing the last expression in the equation of
motion (\ref{mot}) we obtain
    \begin{eqnarray}
        g^{AB}\partial_{A}\partial_{B}\Psi &-&\frac{1}{2}g^{AB}\omega_{(A}^{\,\,\,\,\,\, ab}\sigma_{ab}\partial_{B)}\Psi-
        g^{AB}\omega_{AB}^{\enskip\enskip C}\partial_{C}\Psi+\frac{1}{4}g^{AB}\partial_{A}\omega_{B}^{\enskip ab}\sigma_{ab}\Psi+\nonumber
        \\
        &+&\frac{1}{16}g^{AB}\omega_{A}^{\enskip ab}\omega_{B}^{\enskip cd}\sigma_{ab}\sigma_{cd}\Psi
        -\frac{1}{4}g^{AB}\omega_{AB}^{\enskip\enskip C}\omega_{C}^{\enskip ab}\sigma_{ab}\Psi=0.
    \end{eqnarray}
Here, we have made use of the fact that
$\gamma_{a}\gamma_{b}=\frac{1}{2}\{\gamma_{a},\gamma_{b}\}+\frac{1}{2}[\gamma_{a},\gamma_{b}]=g_{ab}\mathbb{I}+\sigma_{ab}
$, $\omega_{M}^{\enskip ab}=-\omega_{M}^{\enskip ba}$ and
$\omega_{M}^{\enskip ab}\gamma_{a}\gamma_{b}=\omega_{M}^{\enskip
ab}\sigma_{ab}$. Once we simplify some terms, we obtain
    \begin{eqnarray}
        \frac{1}{2}g^{AB}\omega_{(A}^{\,\,\,\,\,\, ab}\sigma_{ab}\partial_{B)}\Psi&=&\frac{1}{2}g^{AB}\omega_{A}^{\enskip ab}\sigma_{ab}\partial_{B}\Psi,\nonumber
        \\
        \omega^{\enskip\enskip C}_{AB}&=&\omega_{A}^{\enskip DC}g_{DB}
        =\omega_{A}^{\enskip dc}\enskip g_{DB} e_{d}^{\enskip D} e_{c}^{\enskip C},\nonumber
        \\
        g^{AB}\omega_{AB}^{\enskip\enskip C}&=&g^{AB}\omega_{A}^{\enskip dc}g_{DB} e_{d}^{\enskip D} e_{c}^{\enskip C}
        =\omega_{A}^{\enskip dc}\delta_{D}^{\enskip A} e_{d}^{\enskip D} e_{c}^{\enskip C} =\omega_{A}^{\enskip dc} e_{d}^{\enskip A} e_{c}^{\enskip C}.\nonumber
    \end{eqnarray}
Finally, the equation of motion for the spinors assumes its final
form
    \begin{eqnarray}
        &&g^{AB}\partial_{A}\partial_{B}\Psi-\frac{1}{2}g^{AB}\omega_{A}^{\enskip ab}\sigma_{ab}\partial_{B}\Psi-
        \omega_{A}^{\enskip ab} e_{a}^{\enskip A} e_{b}^{\enskip C}\partial_{C}\Psi+\frac{1}{4}g^{AB}\partial_{A}\omega_{B}^{\enskip ab}\sigma_{ab}\Psi+\nonumber
        \\
        &&+\frac{1}{16}g^{AB}\omega_{A}^{\enskip ab}\omega_{B}^{\enskip cd}\sigma_{ab}\sigma_{cd}\Psi-
        \frac{1}{4}\omega_{A}^{\enskip ab} e_{a}^{\enskip A} e_{b}^{\enskip C}\omega_{C}^{\enskip
        cd}\sigma_{cd}\Psi=0,\label{emov}
    \end{eqnarray}
which is very difficult to be resolved because the fields are coupled.

\subsection{Conformal Mapping Based Solution}

In order to simplify the structure of the equation (\ref{emov}),
we shall introduce the following transformation on the spinor
components:
    \begin{equation}\nonumber
        \Psi=
        \left(
            \begin{array}{c}
                \varphi_{1} \\
                \varphi_{2} \\
        \end{array}
        \right),
    \end{equation}
where components are grouped as
    \begin{equation}\nonumber
        \varphi_{1}=
        \left(
          \begin{array}{c}
            \psi_{1} \\
            \psi_{2} \\
          \end{array}
        \right)
        ,\enskip
        \varphi_{2}=
        \left(
          \begin{array}{c}
            \psi_{3} \\
            \psi_{4} \\
          \end{array}
        \right).
    \end{equation}
With this representation we obtain the equation of motion for
$\varphi_1$ and $\varphi_2$
    \begin{eqnarray}
        \widehat{{\O}}\varphi_{1}+\frac{3\psi_{0}}{\psi^{2}}\ \frac{\partial\varphi_{1}}{\partial t}-\frac{4}{\psi}\ \frac{\partial
        \varphi_{1}}{\partial \psi}&+&\frac{1}{4\psi^{2}}\varphi_{1}-\frac{i\psi_{0}}{\psi^{2}}\ e^{-\frac{t}{\psi_{0}}}\ \overrightarrow{\sigma}\cdot\overrightarrow{\nabla}
        \varphi_{1}=\nonumber
        \\
        &=&-\frac{i\psi_{0}}{\psi^{2}}\ \frac{\partial\varphi_{2}}{\partial t}+\frac{3i}{2\psi^{2}}\varphi_{2}-\frac{\psi_{0}}{\psi^{2}}\ e^{-\frac{t}{\psi_{0}}}\
        \overrightarrow{\sigma}\cdot\overrightarrow{\nabla}\varphi_{2},\label{sysmot1}
        \\
        \widehat{{\O}}\varphi_{2}+\frac{3\psi_{0}}{\psi^{2}}\ \frac{\partial\varphi_{2}}{\partial t}-\frac{4}{\psi}\ \frac{\partial\varphi_{2}}{\partial \psi}
        &+&\frac{1}{4\psi^{2}}\varphi_{2}+\frac{i\psi_{0}}{\psi^{2}}\ e^{-\frac{t}{\psi_{0}}}\ \overrightarrow{\sigma}\cdot\overrightarrow{\nabla}\varphi_{2}=\nonumber
        \\
        &=&\frac{i\psi_{0}}{\psi^{2}}\ \frac{\partial\varphi_{1}}{\partial t}-\frac{3i}{2\psi^{2}}\varphi_{1}-\frac{\psi_{0}}{\psi^{2}}\ e^{-\frac{t}{\psi_{0}}}\
        \overrightarrow{\sigma}\cdot\overrightarrow{\nabla}\varphi_{1}.\label{sysmot2}
    \end{eqnarray}
Here, we have adopted the following conventions:
    \begin{eqnarray}
        \widehat{{\O}}\varphi&=&\left(\frac{\psi_{0}}{\psi}\right)^{2}\frac{\partial^{2}\varphi}{\partial t^{2}}-\left(\frac{\psi_{0}}{\psi}\right)^{2}
        e^{-\frac{2t}{\psi_{0}}}\ \nabla^{2}\varphi-\frac{\partial^{2}\varphi}{\partial \psi^{2}},\nonumber
        \\
        \overrightarrow{\sigma}&=&\sigma_{1}\ \hat{\imath}+\sigma_{2}\ \hat{\jmath}+\sigma_{3}\ \hat{k},\nonumber
        \\
        \overrightarrow{\sigma}\cdot\overrightarrow{\nabla}\varphi&=&\sigma_{1}\ \frac{\partial\varphi}{\partial x}+\sigma_{2}\ \frac{\partial\varphi}{\partial y}
        +\sigma_{3}\ \frac{\partial\varphi}{\partial z}.\nonumber
    \end{eqnarray}
Now we can use the conformal mapping defining new complex fields
$\Phi_{+}=\varphi_{1}+i\varphi_{2}$ and
$\Phi_{-}=\varphi_{1}-i\varphi_{2}$. Rewriting the equations
(\ref{sysmot1}) and (\ref{sysmot2}) in terms of these new fields,
it is possible to decouple the first equation, while that the
other coupling becomes a source for the second equation
    \begin{eqnarray}
        \widehat{{\O}}\Phi_{+}+\frac{4\psi_{0}}{\psi^{2}}\ \frac{\partial\Phi_{+}}{\partial t}-\frac{4}{\psi}\
        \frac{\partial\Phi_{+}}{\partial
        \psi}-\frac{5}{4\psi^{2}}\Phi_{+}&=&0, \label{homeq}
        \\
        \widehat{{\O}}\Phi_{-}+\frac{2\psi_{0}}{\psi^{2}}\ \frac{\partial\Phi_{-}}{\partial t}-\frac{4}{\psi}\ \frac{\partial\Phi_{-}}{\partial \psi}
        +\frac{7}{4\psi^{2}}\Phi_{-}&=&\frac{i2\psi_{0}}{\psi^{2}}\ e^{-\frac{t}{\psi_{0}}}\
        \overrightarrow{\sigma}\cdot\overrightarrow{\nabla}\Phi_{+}.
        \label{inhomeq}
    \end{eqnarray}
Then, after few calculations, the Lagrangian density written in
terms of the new fields takes the form
    \begin{equation}
        \mathcal{L}_{eff}=-\frac{1}{2}\left(\nabla_{A}\overline{\varphi}_{1}\ \nabla^{A}\varphi_{1}+\nabla_{A}\overline{\varphi}_{2}\ \nabla^{A}\varphi_{2}\right),\nonumber
    \end{equation}
or, alternatively, can be written as a function of the pair $
\varphi_{1}=\frac{1}{2}\left(\Phi_{+}+\Phi_{-}\right),
\varphi_{2}=\frac{1}{2i}\left(\Phi_{+}-\Phi_{-}\right) $
    \begin{equation}\label{lag}
        \mathcal{L}_{eff}=-\frac{1}{4}\left(\nabla_{A}\overline{\Phi}_{+}\ \nabla^{A}\Phi_{+}+\nabla_{A}\overline{\Phi}_{-}\
        \nabla^{A}\Phi_{-}\right).
    \end{equation}
On the other hand the 5D Energy-Momentum (EM) tensor is
represented by
$^{(5)}T_{AB}=2\frac{\delta\mathcal{L}_{eff}}{\delta
g^{AB}}-g_{AB}\mathcal{L}_{eff}$. This procedure take place in a
5D vacuum. Therefore, the effective Lagrangian and the EM tensor
are involved directly with the cosmological observables we wish to
evaluate. The observables to which we refer are energy density and
pressure. Both come from the diagonal part of the EM tensor.

\subsection{Extra dimensional solution for $\Phi_{+}$}

We shall use the variable separation method to the homogeneous PDE
(\ref{homeq}), we obtain the following set of ODE's:

\begin{eqnarray}
        \nabla^{2}R&+&\kappa^{2}R=0, \label{e1}\\
        \frac{\partial^{2}T^{(+)}_{\kappa}}{\partial t^{2}}&+&\frac{2}{\psi_{0}}\frac{\partial T^{(+)}_{\kappa}}{\partial t}+(\kappa^{2}e^{-\frac{2t}{\psi_{0}}}-M_{1}^{2})
        T^{(+)}_{\kappa}=0, \label{e2} \\
        \psi^{2}\frac{\partial^{2}\Lambda}{\partial\psi^{2}}&+&4\psi\frac{\partial\Lambda}{\partial\psi}+(\frac{5}{4}-M_{1}^{2}\psi_{0}^{2})\Lambda=0.
        \label{e3}
\end{eqnarray}
The equation (\ref{e1}) has a solution that can be written in
terms of plane wavefront
\begin{equation}
R(\overrightarrow{r})\sim e^{\pm i \vec{\kappa}\cdot\vec{r}}.
\end{equation}
The second equation (\ref{e2}) has a general solution
\begin{equation}
\Lambda^{(+)}(\psi)=C_{1}\left(\frac{\psi}{\psi_{0}}\right)^{-(\frac{3}{2}+\sqrt{1+M_{1}^{2}\psi_{0}^{2}})}
+C_{2}\left(\frac{\psi}{\psi_{0}}\right)^{-(\frac{3}{2}-\sqrt{1+M_{1}^{2}\psi_{0}^{2}})}.
\end{equation}
Since we are interested in "localized" static solutions, i.e.
those that decay to zero when $\psi$ tends to infinity, we must
choose $C_{2}=0$, so that
$n\equiv {(\frac{3}{2}+\sqrt{1+M_{1}^{2}\psi_{0}^{2}})} > 0$. This choice
makes $M_{1}^{2}=\frac{(n-\frac{3}{2})^{2}-1}{\psi_{0}^{2}}\geq
0$, with $n\geq 3$ and $n \in \mathbb{R}$, in order to
$\sqrt{1+M_{1}^{2}\psi_{0}^{2}} \geq 0$.

\subsection{Extra dimensional solution for $\Phi_{-}$}

Now we are able to calculate the coupling term of the
inhomogeneous equation (\ref{inhomeq}) for each mode [see eq. (\ref{treinta})]
\begin{eqnarray}
\frac{2\,i\,\psi_{0}}{\psi^{2}}\ e^{-\frac{t}{\psi_{0}}}\
\vec{\sigma}\cdot\vec{\nabla}\Phi_{+,\kappa}&=&-\frac{2^{\nu}
\Gamma(\nu)}{\sqrt{\pi}\,\kappa^{\nu-1} \,\psi_0^{\frac{1}{2}
+\nu}} \,
e^{i\,\vec{\kappa}\cdot\vec{x}}\,e^{(\nu-3)\,\frac{t}{\psi_{0}}}\enskip
\,\left(\frac{\psi}{\psi_{0}}\right)^{-(\frac{7}{2}+\sqrt{\nu^2-3})}.
\end{eqnarray}
Using the last expression in eq. (\ref{inhomeq}), we obtain a
degenerate two-component system\footnote{Henceforth we are
concerned with asymptotic solutions, i.e. only the infrared limit
makes cosmological significance.} for the spinor
$\Phi_{-,\kappa}$. Again, a plane wavefront satisfies the spatial
part. By inserting
$\Phi_{-,\kappa}=G_{\kappa}(t,\psi)e^{i\vec{\kappa}\cdot\vec{r}}$,
and multiplying by $\left(\frac{\psi}{\psi_{0}}\right)^{2}$, we
obtain
\begin{eqnarray}
\frac{\partial^{2}G^{(-)}_{\kappa}}{\partial
t^{2}}+\frac{2}{\psi_{0}}\frac{\partial G^{(-)}_{\kappa}}{\partial
t}&+&\left(\kappa^{2}
e^{-\frac{2t}{\psi_{0}}}+\frac{7}{4\psi_{0}^{2}}\right)G_{\kappa}-\left[\left(\frac{\psi}{\psi_{0}}\right)^{2}\frac{\partial^{2}G^{(-)}_{\kappa}}{\partial
\psi^{2}}+\frac{4\psi}{\psi_{0}^{2}}\frac{\partial
G^{(-)}_{\kappa}}{\partial \psi}\right]\simeq\nonumber
\\
&\simeq& -\frac{2^{\nu} \Gamma(\nu)\,\kappa^{1-\nu}}{\sqrt{\pi}
\psi_0^{\frac{1}{2}+\nu}}\enskip
e^{\frac{\nu-3}{\psi_{0}}t}\enskip
\left(\frac{\psi}{\psi_{0}}\right)^{-\left[\frac{3}{2}+\sqrt{\nu^2-3}\right]},
\label{uu}
\end{eqnarray}
This inhomogeneous PDE can be converted to one with a constant
coupling. In order to make constant the right side of eq.
(\ref{uu}), we shall propose
\begin{equation}\nonumber
G^{(-)}_{\kappa}(t,\psi)= \psi^{-2}_0\,\,e^{\frac{\nu-3}{\psi_{0}}t}\enskip
\left(\frac{\psi}{\psi_{0}}\right)^{-\left[\frac{3}{2}+\sqrt{\nu^2-3}\right]}\enskip
K^{(-)}_{\kappa}(t,\psi).
\end{equation}
Finally, we must solve the equation
\begin{eqnarray}
\frac{\partial ^{2}K^{(-)}_{\kappa}}{\partial
t^{2}}&+&\frac{2(\nu-2)}{\psi_{0}}\frac{\partial
K^{(-)}_{\kappa}}{\partial t}-\left(\frac{\psi}{\psi_0}\right)^2
\frac{\partial K^{(-)}_{\kappa}}{\partial \psi^2} -
\frac{2}{\psi^2_0} \left[\frac{1}{2} - \sqrt{\nu^2-3}\right] \psi
\frac{\partial K^{(-)}_{\kappa}}{\partial\psi} \nonumber \\
&+& \left[\kappa^2 e^{-2t/\psi_0} + \frac{10-4\nu}{\psi^2_0}
\right] K^{(-)}_{\kappa}\simeq - \frac{2^{\nu}
\Gamma(\nu)}{\sqrt{\pi} \kappa^{\nu-1} \psi_0^{-\frac{3}{2}+\nu}}.
\label{...}
\end{eqnarray}

\section{Effective dynamics on the 4D hypersurface $\psi=1/H_0$}

In order to describe the effective 4D dynamics of the physical
system in the early inflationary universe with an effective 4D de
Sitter expansion, we shall consider a static foliation on the 5D
metric (\ref{met1}). The resulting 4D hypersurface after making
the static foliation $\psi=\psi_0=1/H_0$, describes an effective 3D spatially flat, isotropic and homogeneous
de Sitter four-dimensional expanding universe with a constant Hubble parameter $H_{0}$, with a line element
\begin{equation}\label{met2}
dS^{2}\rightarrow ds^{2}=dt^{2}-e^{2H_{0}t}dr^{2},
\end{equation}
From the
relativistic point of view, an observer who moves in a co-moving
frame with the five-velocity $U^\psi=0$ on a 4D hypersurface with
a scalar curvature $^{(4)} R=12/\psi^2_0=12\,H^2_0$, such that the
Hubble parameter $H_0$, and thus also the cosmological constant:
$\Lambda_0=3H^2_0/(8\pi G)$, are defined by the foliation
$H_0=\psi_0^{-1}$.

\subsection{Time dependent modes of $\Phi_{+}$}

The solution for the
time-dependent equation (\ref{e3}) can be expanded in terms of
first and second kind Hankel functions
\begin{equation}
T^{(+)}_{\kappa}(t)=e^{-2 H_0 t} \left[ C_{3} \, {\cal
H}^{(1)}_{\nu}\left(\frac{\kappa}{H_0}\,e^{- H_0 t}\right)+C_{4}\,{\cal
H}^{(2)}_{\nu}\left(\frac{\kappa}{H_0}\,e^{- H_0 t}\right)
\right],
\end{equation}
where $\nu= \sqrt{4+M^2_1 \psi^2_0}\geq 2$. After make a
Bunch-Davies normalization of the modes\cite{bd} we obtain the
solution
\begin{equation}
T^{(+)}_{\kappa}(t) = \frac{i}{2}\, \sqrt{\frac{\pi}{H_0}}\,\,
e^{-2 H_0 t}\,\,{\cal
H}^{(2)}_{\nu}\left(\frac{\kappa}{H_0}\,e^{- H_0 t}\right).
\end{equation}
Since we are interested to describe the universe on super-Hubble
cosmological scales we must require
$\kappa\,\psi_{0}\,e^{-\frac{t}{\psi_{0}}}\ll 1$, we reject
solutions that goes to zero at late times. The asymptotic behavior
of $T^{(+)}_{\kappa}(t)$ on cosmological scales will be
\begin{equation}
T^{(+)}_{\kappa} (t)\simeq \frac{i}{2}\,\, \sqrt{\frac{\pi}{H_0}}\,\,\Gamma(\nu)\,
e^{-2 H_0 t}\,\,\left(\frac{\kappa}{2 H_0}\,e^{- H_0 t}\right)^{-\nu}.
\end{equation}
Finally, the degenerate two-component spinor $\Phi_{+}$ can be
expanded as a function of the modes
\begin{equation}
\Phi_{{+},\kappa}(t,\vec{r},\psi_0=1/H_0)\simeq i\, C_1\,\frac{2^{\nu-1}
\Gamma(\nu)}{\sqrt{\pi}} \,H_0^{\nu-\frac{1}{2}}
\,\kappa^{-\nu}\enskip e^{i \vec{\kappa}\cdot\vec{r}}\enskip
e^{(\nu-2)\,H_0 t},
\end{equation}\label{treinta}
and their complex conjugated.

\subsection{The time dependent modes for $\Phi_{-}$}

The homogeneous solution $\left.K^{(-)}_{\kappa}(t,\psi)\right|_{hom}$, of the eq. (\ref{...}), is
\begin{eqnarray}
\left.K^{(-)}_{\kappa}(t,\psi)\right|_{hom}& = & e^{-\frac{(\nu-2)t}{\psi_{0}}} \left[ \bar{C}_{3} \, {\cal
H}^{(1)}_{\mu}\left(\kappa\,\psi_{0}\,e^{-\frac{t}{\psi_{0}}}\right)+\bar{C}_{4}\,{\cal
H}^{(2)}_{\mu}\left(\kappa\,\psi_{0}\,e^{-\frac{t}{\psi_{0}}}\right)
\right] \nonumber \\
& \times &
\left[\bar{C}_{1}\left(\frac{\psi}{\psi_{0}}\right)^{(\sqrt{\nu^2-3} - \sqrt{\nu^2-4\nu +7+M^2_2\psi^2_0})}
+\bar{C}_{2}\left(\frac{\psi}{\psi_{0}}\right)^{(\sqrt{\nu^2-3} + \sqrt{\nu^2-4\nu +7+M^2_2\psi^2_0})}\right]. \nonumber \\
\end{eqnarray}
where $\mu= \sqrt{\nu^2-4\nu +4+M^2_2\psi^2_0} \geq 0$ and and the squared mass of $\Phi_{(-)}$ is
$[M_2(m,n)]^2 = {(m-3/2)^2-(n-3/2)^2+4\sqrt{(n-3/2)^2+3}-10\over \psi^2_0}$, which is definite positive for $m\geq 1$ (with $n\geq 3$).
In order to the $\psi$-dependent solution of
$\lim_{\psi\rightarrow \infty} {\left.K^{(-)}_{\kappa}(t,\psi)\right|_{hom}}\rightarrow 0$, we shall require that
$\bar{C}_{2}=0$ and $m> 3/2+\sqrt{3+\left(\sqrt{(n-3/2)^2+3}-2\right)^2}$, for $n\geq 3$, such that
$(m,n)\in \mathbb{Z}$.
After take the asymptotic limit on cosmological scales we obtain that the modes $\Phi_{-,\kappa}(t,\vec{r},\psi_0=1/H_0)$, for $\mu=1$, are
\begin{equation}
\Phi_{-,\kappa}(t,\vec{r},\psi_0) \simeq A_2 \, \frac{H^{1/2}_0}{\sqrt{\pi} \kappa} \,e^{i \vec{\kappa}.\vec{r}},
\end{equation}
where $A_2 = \bar{C}_4 \bar[C]_1$. Notice that we have neglected the inhomogenoues part of its solution because it is negligible on these
large super-Hubble scales at the end of inflation.

As can be demonstrated the solution $K^{(-)}_{\kappa}(t,\psi_0)
\simeq \left.K^{(-)}_{\kappa}(t,\psi_0)\right|_{hom}$ on
cosmological scales, once we consider $H_0=1/\psi_0 = 1\times
10^{-9}\, {\rm M_p}$. Hence, the homogeneous solution
$\left.K^{(-)}_{\kappa}(t,\psi_0)\right|_{hom}$ is a very
acceptable solution at the end of inflation for the time dependent
modes for the time dependent modes of $\Phi_{-}$. In other words,
at the end of inflation the effective 4D bosons $\Phi_{\pm}$ can
be decoupled on cosmological scales.

\subsection{4D Einstein equations}

The effective 4D Lagrangian density (\ref{lag}) is expressed in
terms of the fields $\Phi_{\pm}(x^{\mu},\psi_0)$, which can be
thought of as two minimally coupled bosons
\begin{equation}
\mathcal{L}_{eff}=-\frac{1}{4}\left[\nabla_{\mu}\overline{\Phi}_{+}\
\nabla^{\mu}\Phi_{+}+\nabla_{\mu}\overline{\Phi}_{-}\,\nabla^{\mu}\Phi_{-}\right
]+ V\left(\Phi_{+},\Phi_{-}\right).
\end{equation}
Since
\begin{eqnarray}
\nabla_4 \bar\Phi_{+} &= &{\partial \bar\Phi_{+}\over
\partial\psi} \left( 1 \,\, 1\right)
=-(n/\psi)\,\bar\Phi_{+}\, \left( 1 \,\, 1\right), \\
\nabla_4 \bar\Phi_{-} &=& {\partial \bar\Phi_{-}\over
\partial\psi} \left(1 \,\, 1\right)=-(m/\psi)\,\bar\Phi_{-}\,\left( 1 \,\, 1\right), \\
\nabla_4 {\Phi}_{+} &= &{\partial {\Phi}_{+}\over
\partial\psi} \left(\begin{array}{ll} 1 \\ 1 \end{array} \right)
=-(n/\psi)\,{\Phi}_{+}  \left(\begin{array}{ll} 1 \\ 1 \end{array} \right), \\
\nabla_4 {\Phi}_{-} &= &{\partial {\Phi}_{-}\over
\partial\psi} \left(\begin{array}{ll} 1 \\ 1 \end{array} \right)=-(m/\psi)\,{\Phi}_{-} \left(\begin{array}{ll} 1 \\ 1 \end{array} \right),
\end{eqnarray}
hence the effective 4D potential results to be
\begin{eqnarray}
V\left(\Phi_{+},\Phi_{-}\right)&=&-\left.{1\over 4}
\left[\nabla_{4}\overline{\Phi}_{+}\
\nabla^{4}\Phi_{+}+\nabla_{4}\overline{\Phi}_{-}\
\nabla^{4}\Phi_{-}\right]\right|_{\psi=1/H_0}\nonumber \\
&=& \left.{H^2_0\over 4}
\left(n^2 \|\Phi_{+}\|^2 + m^2
\|\Phi_{-}\|^2\right)\right|_{\psi=1/H_0},
\end{eqnarray}
which is induced by the static foliation on the fifth coordinate
$\psi=\psi_0=1/H_0$. This effective 4D potential is the responsible to provide us the dynamics
of the fields $\Phi_{\pm}(x^{\mu},\psi_0)$ on the effective 4D
hypersurface on which the equation of state is $\omega=P/\rho=-1$.
The energy density and pressure related to these fields are obtained
from the diagonal part of the energy-momentum tensor written in a
mixed manner
\begin{eqnarray}
\rho &=& \left<E\left| \frac{1}{4} \left[\left\|\nabla_0
\Phi_+\right\|^2 + \left\|\nabla_0 \Phi_-\right\|^2\right] -
\frac{e^{-2 H_0 t}}{4} \left[ \vec{\nabla}\Phi_{-} .
 \vec{\nabla} \bar\Phi_{-} + \vec{\nabla}\Phi_{+} .\vec{\nabla}\bar{\Phi}_{+}\right]
\right. \right.\nonumber \\
& + &\left.\left.\left.
V\left(\Phi_+,\Phi_-\right)+F^0_{\,\,0}\right|E\right>
\right|_{\psi=1/H_0},
\\
 P &=& \left<E\left| \frac{1}{4} \left[\left\|\nabla_0
\Phi_+\right\|^2 + \left\|\nabla_0 \Phi_-\right\|^2\right] -
\frac{e^{-2 H_0 t}}{12} \left[ \vec{\nabla}\Phi_{-} .
 \vec{\nabla} \bar\Phi_{-} + \vec{\nabla}\Phi_{+} .\vec{\nabla}\bar{\Phi}_{+}\right]
\right.\right. \nonumber \\
& -& \left.\left. \left.V\left(\Phi_+,\Phi_-\right)+F^i_{\,\,j}
\delta^i_j\right|E\right>\right|_{\psi=1/H_0},
\end{eqnarray}
where $\left.|E\right>$ is some quantum
state, $F^0_{\,\,\,0} = C_3/\pi \left[{H^7_0\over 8\kappa^2}+{H^9_0\over
\kappa^4}\right]$, $F^i_{\,\,\,j} = A_3/\pi \left[{15 H^7_0\over 32
\kappa^2}+{H^9_0 \over 2 \kappa^4}\right] \delta^i_j$ and
\begin{eqnarray}
\nabla_0 \Phi_{\pm} &= & \left[\partial_0  \mp {1\over 4\psi_0}\right] \Phi_{\pm}\left(\begin{array}{ll} 1 \\ 1 \end{array} \right), \\
\nabla_j \Phi_{+} &= & \partial_j  \Phi_{+}\left(\begin{array}{ll} 1 \\ 1 \end{array} \right), \\
\nabla_1 \Phi_{-} &=& \left[\partial_1 \Phi_{-} + i\, \frac{H_0 e^{H_0 t}}{2} \Phi_{-} \right] \left(\begin{array}{ll} 1 \\ 1 \end{array} \right), \\
\nabla_2 \Phi_{-} &=& \partial_2 \Phi_{-} \left(\begin{array}{ll} 1 \\ 1 \end{array} \right) + i\, \frac{H_0 e^{H_0 t}}{2} \Phi_{-}  \left(\begin{array}{ll} -i \\ i
\end{array} \right), \\
\nabla_3 \Phi_{-} &=& \partial_3 \Phi^*_{-} \left(\begin{array}{ll} 1 \\ 1 \end{array} \right) + i\, \frac{H_0 e^{H_0 t}}{2} \Phi^*_{-}  \left(\begin{array}{ll} 1 \\ -1
\end{array} \right), \\
\nabla_1 \bar{\Phi}_{-} &=& \left[\partial_1 {\Phi}_{-} - i\, \frac{H_0 e^{H_0 t}}{2} {\Phi}_{-} \right] \left( 1 \,\, 1 \right), \\
\nabla_2 \bar{\Phi}_{-} &=& \partial_2 {\Phi}^*_{-} \,\left( 1 \,\, 1 \right) - i\, \frac{H_0 e^{H_0 t}}{2} {\Phi}^*_{-}  \left( i \,\, -i \right), \\
\nabla_3 \bar{\Phi}_{-} &=& \partial_3 {\Phi}^*_{-} \,\left( 1 \,\, 1 \right) - i\, \frac{H_0 e^{H_0 t}}{2} {\Phi}^*_{-}  \left( 1 \,\, -1 \right).
\end{eqnarray}
An interesting asymptotic solution can be obtained by considering the expectation values of, for instance, some quadratic scalar
$\Sigma^2(\vec{x},t)$, as
\begin{equation}
\Sigma^2(t)=\left<E\left|\Sigma^2(\vec{x},t)\right|E\right> = \frac{1}{(2\pi)^3 }\int^{\epsilon \kappa^{\pm}_0(t)}_{\kappa_*} d^3\kappa \,\,\Sigma_\kappa(\vec{x},t)
\Sigma^*_\kappa(\vec{x},t),
\end{equation}
where $\kappa_*>0$ is some minimum cut for the wavenumber to be
determined and $\kappa^{+}_0(t)= H_0 e^{H_0 t}$, $\kappa^{-}_0(t)= 2 H_0
e^{H_0 t}$ are the maximum wavenumbers to the modes of $\Phi_{+}$
and $\Phi_{-}$, respectively. The expectation values for the
radiation energy density $\rho$ and the pressure $P$, are given by
the expressions
\begin{eqnarray}
\rho &=& \left[A_2^2\left(\frac{173H_0^4\epsilon}{128\pi^3}-\frac{\epsilon^3H_0^4}{24\pi^3}\right)+C_3\frac{H_0^8\epsilon}{16\pi^3}\right] e^{H_0 t}-A_2^2\frac{173k_*H_0^3}{128\pi^3}
+C_1^2\frac{101H_0^5}{32k_*\pi^3}+ \nonumber\\
&&+C_3\left(-\frac{k_*H_0^7}{16\pi^3}+\frac{H_0^9}{2k_*\pi^3}\right)-\left[C_1^2\left(\frac{101H_0^4}{64\pi^3\epsilon}+\frac{H_0^4\epsilon}{\pi^3}\right)
+C_3\frac{H_0^8}{4\pi^3\epsilon}\right] e^{-H_0 t}+ \nonumber\\
&&+\left[C_1^2\frac{k_*H_0^3}{2\pi^3}+A_2^2\frac{k_*^3H_0}{24\pi^3}\right] e^{-2H_0 t}, \\
P &=&
\left[A_3\frac{15H_0^8\epsilon}{64\pi^3}-A_2^2\left(\frac{219H_0^4\epsilon}{128\pi^3}+\frac{H_0^4\epsilon^3}{24\pi^3}\right)\right]
e^{H_0 t}
-C_1^2\frac{99H_0^5}{32k_*\pi^3}+A_2^2\frac{219k_*H_0^3}{128\pi^3}+\nonumber \\
&&+A_3\left(-\frac{15k_*H_0^7}{64\pi^3}+\frac{H_0^9}{4k_*\pi^3}\right)+\left[C_1^2\left(\frac{99H_0^4}{64\pi^3\epsilon}-\frac{H_0^4\epsilon}{\pi^3}\right)
-A_3\frac{H_0^8}{8\pi^3\epsilon}\right] e^{-H_0 t}+\nonumber \\
&&+\left[A_2^2\frac{k_*^3H_0}{24\pi^3}+C_1^2\frac{k_*H_0^3}{2\pi^3}\right]
e^{-2H_0 t}.
\end{eqnarray}
Since we are interested to find solutions with $\mu=1$ and $\nu=2$ that correspond to $\partial_0 \Phi_{\pm}=0$, we must consider
the values $n=5/2$, $m=7/2$. In order to cancelate the coefficients corresponding to the factors $e^{\pm H_0t}$ and if we require that
$\rho=-P = 3H^2_0/(8\pi G)$, we obtain that
\begin{small}
\begin{eqnarray}
C^2_1 &=&
\frac{6k_*\pi^2(-519+16\epsilon^2)}{H_0\epsilon^2[32H_0^2(-519+16\epsilon^2)-k_*^2(101+64\epsilon^2)]}
\nonumber \\
&=&
\frac{12k_*\pi^2(657+16\epsilon^2)}{H_0\epsilon^2[64H_0^2(657+16\epsilon^2)-15k_*^2(-99+64\epsilon^2)]},
\\
A^2_2 &=&
-\frac{3C_1^2(101+64\epsilon^2)}{2(-519+16\epsilon^2)}=-\frac{45C_1^2(-99+64\epsilon^2)}{4(657+16\epsilon^2)},
\label{48}
\end{eqnarray}
\end{small}
such that from eq. (\ref{48}) we obtain that $\epsilon = 6.68586$.
Furthermore, due to the fact the equation of state is
$\rho=-P=3H^2_0/(8\pi G)$, we must require that
\begin{small}
\begin{eqnarray}
\frac{32H_0^2(519-16\epsilon^2)+k_*^2(101+64\epsilon^2)}{k_*(-519+16\epsilon^2)}
=\frac{64H_0^2(657+16\epsilon^2)-15k_*^2(-99+64\epsilon^2)}{2k_*(657+16\epsilon^2)},
\nonumber
\end{eqnarray}
\end{small}
from which we obtain that $\kappa_* = 1.45598 \, H_0= 1.45598
\times 10^{-9} \,{\rm M_p}$. Notice that we have neglected in $P$
and $\rho$ terms which are very small with respect $3H^2_0/(8\pi
G)$ and decrease as $e^{-2H_0 t}$. With the values earlier
mentioned for $\kappa_*$, $H_0$ and $M_p$, we arrive at the
numerical values $(C_1)^2 = -2.63042 \times 10^{32}$, $(A_2)^2
=5.95601 \times 10^{33}$, $C_3 =4.8693\times 10^{70} {\rm
M^{-4}_p}$, $A_3 =9.081\times 10^{70} {\rm M^{-4}_p}$, that
correspond to $\rho = -P= 1.19366 \times 10^{-19} \,{\rm M^4_p}$.
These values are perfectly according to which one expects during a
inflationary vacuum dominated expansion of the early universe. A
very important fact is that the dark energy is outside the horizon
at the beginning of inflation, but during the inflationary epoch
enters to causally connected regions. In other word the dark
energy is concentrated on the range of scales (physical scales) $
 2\pi/[\epsilon\kappa^{\pm}_0(t)]\simeq (\pi/H_0) e^{-H_0 t} <
\lambda_{phys} < 2\pi/\kappa_*$. Hence, the effective 4D
scalar (massive) field $\Phi_{-}$ should be an interesting
candidate to explain dark energy in the early inflationary
universe.

\section{Final Remarks}

We have explored the possibility that the expansion of the
universe during the primordial inflationary phase of the universe
can be driven by a condensate of spinor fields. In our picture
$\phi_{\pm}$ are effective fields which became from a condensate
of two entangled spinors. The fields $\phi_{\pm}$ decouple at the
end of inflation. In all our analysis we have neglected the role
of the inflaton field, which (in a de Sitter expansion) is freezed
in amplitude and nearly scale invariant, but decays at the end of
inflation into other fields. The point here is how we explain the
existence of dark energy once the inflaton field energy density
goes to zero. Our proposal consist to prove that the dark energy
could be physically explained though the entanglement of spinor
fields that behave as effective 1-spin and 0-spin bosons on a 4D
hypersurface on which the universe suffers a vacuum dominated
expansion. The equation of state of the universe is determined by
the static foliation $\psi=1/H_0$. Our calculations show that the
vector boson $\phi_{+}$ is massless and with spin $1$, and
therefore compatible with the properties of a massless vector
boson. On the other hand the field $\phi_{-}$ is a scalar boson
which could be (jointly with the inflaton) the responsible for the
expansion of the universe and would be a good candidate to explain
the existence of the dark energy. [Other fields such as the
curvaton field\cite{M}, have been proposed in the literature to
explain it.] A very interesting fact is that the (dark) energy
density which we are talking about is poured into smaller
sub-horizon scales with the evolution of the inflationary
expansion.

\section*{Acknowledgements}

\noindent The authors acknowledge UNMdP and CONICET Argentina for
financial support.

\end{document}